\documentclass[twocolumn]{aastex63}
\usepackage{newtxtext,newtxmath}

\renewcommand{\Angstrom}{\text{\ensuremath{\mathring{\mathrm{A}}}}}

\hypersetup{linkcolor=teal, citecolor=teal, filecolor=teal, urlcolor=teal}

\newcommand{\oiii}{[\ion{O}{3}]}
\newcommand{\feii}{\ion{Fe}{2}}

\newcommand{\hei}{\ion{He}{1}}
\newcommand{\nii}{[\ion{N}{2}]}

\newcommand{\ha}{H\ensuremath{\alpha}}
\newcommand{\hb}{H\ensuremath{\beta}}

\def\lsim{\mathrel{\rlap{\lower 3pt \hbox{$\sim$}} \raise 2.0pt \hbox{$<$}}}
\def\gsim{\mathrel{\rlap{\lower 3pt \hbox{$\sim$}} \raise 2.0pt \hbox{$>$}}}

\def\M1450{{\rm M}_{\rm 1450}}

\usepackage{graphicx}
\usepackage{natbib}
\usepackage{relsize}
\usepackage{textcomp}
\usepackage{upgreek}
\usepackage{amsmath}
\usepackage{amssymb}
\usepackage{calrsfs}
\usepackage[switch]{lineno} 
\usepackage{afterpage}
\usepackage{float}
\usepackage{makecell}
\usepackage{multirow}

\usepackage{iftex}

\newcommand{\CJKchars}[1]{#1}

\ifPDFTeX
  \usepackage{CJKutf8}
  \renewcommand{\CJKchars}[1]{{\begin{CJK*}{UTF8}{gbsn}#1\end{CJK*}}}
\fi

\ifXeTeX
  \usepackage{fontspec}
  \usepackage{xeCJK}
\fi

\ifLuaTeX
  \usepackage{fontspec}
  \usepackage{luatexja-fontspec} 
\fi

\setcitestyle{notesep={}}

\received{01 09 2026}
\revised{21 09 2026}
\accepted{03 10 2026}
\published{10 11 2026} 
\submitjournal{ApJL}

\defcitealias{Borisova2016}{B16}
\defcitealias{Cai2019}{C19}
\defcitealias{Arrigoni2019}{AB19}
\defcitealias{Farina2019}{F19}

\shorttitle{A new quasar pair at $z\sim6.08$}

\shortauthors{Onorato et al.}

\graphicspath{{./}{figures/}}
\begin{document}

\title{\textit{N\=ap\=owaw\=a`enakaulua}: a close quasar pair at cosmic dawn in the Aether survey}

\correspondingauthor{Silvia Onorato}
\email{silvia.onorato@noirlab.edu}

\author[0009-0009-1715-4157]{Silvia Onorato}
\affiliation{International Gemini Observatory, NSF NOIRLab, 670 N A`ohoku Place, HI-96720, Hilo, USA}

\author[0000-0002-6822-2254]{Emanuele Paolo Farina}
\affiliation{International Gemini Observatory, NSF NOIRLab, 670 N A`ohoku Place, HI-96720, Hilo, USA}
\affiliation{INAF -- Osservatorio di Astrofisica e Scienza dello Spazio di Bologna, via Gobetti 93/3, I 40129, Bologna, Italy}

\author[0000-0002-2662-8803]{Roberto Decarli}
\affiliation{INAF -- Osservatorio di Astrofisica e Scienza dello Spazio di Bologna, via Gobetti 93/3, I 40129, Bologna, Italy}

\author[0009-0008-2205-7725]{Klaudia Protu\v{s}ov{\'a}}
\affiliation{Institute for Theoretical Physics, Heidelberg University, Philosophenweg 12, D–69120, Heidelberg, Germany}

\author[0000-0002-9712-0038]{Elia Pizzati}
\affiliation{Center for Astrophysics | Harvard \& Smithsonian, 60 Garden St., Cambridge, MA 02138, USA}

\author[0000-0002-3007-0013]{Debora Pelliccia}
\affiliation{UCO/Lick Observatory, Department of Astronomy \& Astrophysics, UC Santa Cruz, 1156 High Street, Santa Cruz, CA 95064, USA}

\author[0000-0003-3765-8001]{Devin S.\ Chu}
\affiliation{`Imiloa Astronomy Center of Hawai`i, University of Hawai`i - Hilo, 600 `Imiloa Place, HI-96720, Hilo, USA}

\author{Larry Lindsey Kimura}
\affiliation{Ka Haka `Ula O Ke`elik\=olani/College of Hawaiian Language (KHUOK)}

\author{Leinani Lozi}
\affiliation{International Gemini Observatory, NSF NOIRLab, 670 N A`ohoku Place, HI-96720, Hilo, USA}

\author{Emily Peavy}
\affiliation{International Gemini Observatory, NSF NOIRLab, 670 N A`ohoku Place, HI-96720, Hilo, USA}

\author[0000-0002-4770-6137]{Fabrizio Arrigoni--Battaia}
\affiliation{Max Planck Institut f\"ur Astrophysik, Karl--Schwarzschild--Stra{\ss}e 1, D-85748, Garching bei M\"unchen, Germany}

\author[0000-0002-2931-7824]{Eduardo Ba\~{n}ados}
\affiliation{Max Planck Institut f\"ur Astronomie, K\"onigstuhl 17, D-69117 Heidelberg, Germany}

\author[0000-0002-3026-0562]{Aaron J.\ Barth}
\affiliation{Department of Physics and Astronomy, 4129 Frederick Reines Hall, University of California, Irvine, CA, 92697-4575, USA}

\author[0000-0003-4747-4484]{Silvia Belladitta}
\affiliation{Max Planck Institut f\"ur Astronomie, K\"onigstuhl 17, D-69117 Heidelberg, Germany}
\affiliation{INAF -- Osservatorio di Astrofisica e Scienza dello Spazio di Bologna, via Gobetti 93/3, I 40129, Bologna, Italy}

\author[0000-0002-4314-021X]{Manuela Bischetti}
\affiliation{Dipartimento di Fisica ”Enrico Fermi”, Università di Pisa, Largo Bruno Pontecorvo 3, Pisa, I-56127, Italy}
\affiliation{INAF - Osservatorio Astronomico di Trieste, Via G. B. Tiepolo 11, I-34143 Trieste, Italy}

\author[0000-0001-8582-7012]{Sarah E.~I.~Bosman}
\affiliation{Institute for Theoretical Physics, Heidelberg University, Philosophenweg 12, D–69120, Heidelberg, Germany}
\affiliation{Max Planck Institut f\"ur Astronomie, K\"onigstuhl 17, D-69117 Heidelberg, Germany}

\author[0000-0002-3173-1098]{Hyunseop Choi}
\affiliation{Department of Astronomy, University of Michigan, 1085 S. University Ave., Ann Arbor, MI 48109, USA}

\author[0000-0002-6748-2900]{Tiago Costa}
\affiliation{School of Mathematics, Statistics and Physics, Newcastle University, Newcastle upon Tyne, NE1 7RU, UK}

\author[0000-0003-2895-6218]{Anna-Christina Eilers}
\affiliation{Department of Physics, Massachusetts Institute of Technology, Cambridge, MA 02139, USA}
\affiliation{MIT Kavli Institute for Astrophysics and Space Research, Massachusetts Institute of Technology, Cambridge, MA 02139, USA}

\author[0000-0003-3310-0131]{Xiaohui Fan}
\affiliation{Steward Observatory, University of Arizona, 933 N. Cherry Ave., Tucson, AZ 85721, USA}

\author[0000-0002-7200-8293]{Simona Gallerani}
\affiliation{Scuola Normale Superiore, Piazza dei Cavalieri 7, I-56126 Pisa, Italy}

\author[0000-0003-2824-3875]{Thomas R.\ Geballe}
\affiliation{International Gemini Observatory, NSF NOIRLab, 670 N A`ohoku Place, HI-96720, Hilo, USA}

\author[0009-0009-8274-441X]{Anniek J.\ Gloudemans}
\affiliation{NSF NOIRLab, Gemini Observatory, 670 N A'ohoku Place, Hilo, HI 96720, USA}

\author[0000-0002-3370-187X]{Eunchong Kim}
\affiliation{International Gemini Observatory, NSF NOIRLab, 670 N A`ohoku Place, HI-96720, Hilo, USA}

\author[0000-0002-1428-7036]{Brian C.\ Lemaux}
\affiliation{International Gemini Observatory, NSF NOIRLab, 670 N A`ohoku Place, HI-96720, Hilo, USA}

\author[0000-0003-3762-7344]{Weizhe Liu (\CJKchars{刘伟哲})}
\affiliation{Steward Observatory, University of Arizona, 933 N. Cherry Ave., Tucson, AZ 85721, USA}

\author[0000-0001-5063-0340]{Yoshiki Matsuoka}
\affiliation{Research Center for Space and Cosmic Evolution, Ehime University, Matsuyama, Ehime 790-8577, Japan}

\author[0000-0002-5941-5214]{Chiara Mazzucchelli}
\affiliation{Instituto de Estudios Astrof\'{\i}sicos, Facultad de Ingenier\'{\i}a y Ciencias, Universidad Diego Portales, Avenida Ejercito Libertador 441, Santiago, Chile}

\author[0009-0008-8066-0717]{Atsuko Nitta}
\affiliation{International Gemini Observatory, NSF NOIRLab, 670 N A`ohoku Place, HI-96720, Hilo, USA}

\author[0000-0002-4544-8242]{Jan-Torge Schindler}
\affiliation{Hamburger Sternwarte, University of Hamburg, Gojenbergsweg 112, D-21029 Hamburg, Germany}

\author[0000-0002-4434-2307]{Andrew W.\ Stephens}
\affiliation{International Gemini Observatory, NSF NOIRLab, 670 N A`ohoku Place, HI-96720, Hilo, USA}

\author[0000-0002-2536-1633]{Hyewon Suh}
\affiliation{International Gemini Observatory, NSF NOIRLab, 670 N A`ohoku Place, HI-96720, Hilo, USA}

\author[0000-0003-4793-7880]{Fabian Walter}
\affiliation{Max Planck Institut f\"ur Astronomie, K\"onigstuhl 17, D-69117 Heidelberg, Germany}

\author[0000-0002-7633-431X]{Feige Wang}
\affiliation{Department of Astronomy, University of Michigan, 1085 S. University Ave., Ann Arbor, MI 48109, USA}

\author[0000-0001-5287-4242]{Jinyi Yang}
\affiliation{Department of Astronomy, University of Michigan, 1085 S. University Ave., Ann Arbor, MI 48109, USA}

\begin{abstract}
We present the discovery and first characterization of \textit{N\=ap\=owaw\=a`enakaulua}: the most compact known quasar pair candidate in the Epoch of Reionization, with a projected separation of only $0.98^{\prime\prime}$ ($\approx 5.56$ pkpc at $z \sim 6.08$) and a velocity offset of $\Delta v \simeq 15 \text{ km s}^{-1}$.
Using spectroscopy from JWST/NIRSpec integral field unit, as part of the \textit{Aether survey}, Gemini/GNIRS and GMOS we characterize the continuum and emission line properties of the two quasars, enabling constraints on their black hole masses ($M_{\text{BH}} \simeq 1.5 \times 10^9$ and $4.0 \times 10^8\,M_\odot$), accretion states ($\lambda_{\text{Edd}} \simeq 1.23$ and $1.07$), and systemic redshifts ($z_{\rm sys} \simeq 6.0781$ and $6.0785$).
We place this pair in the cosmological context by comparing the chance of occurrence to extrapolations of high-$z$ quasar clustering measurements at larger scales, resulting in a $\sim 2-3$ order of magnitude excess with respect to expectations.
This excess hints at an underlying population of merger-triggered quasar pairs at cosmic dawn or to a denser-than-average galactic environment.
\textit{N\=ap\=owaw\=a`enakaulua} thus represents a crucial signpost to investigate the mass assembly of the first massive galaxies and black holes.
\end{abstract}


\section{Introduction}\label{sec:introduction}
The first billion years of cosmic time were a period of furious assembly; dark matter halos collapsed, galaxies ignited their first generations of stars, and the very first supermassive black holes (SMBHs) grew to $10^8-10^9 \,\mathrm{M}_\odot$ \citep[e.g.,][]{Volonteri2012, Inayoshi2020, Fan2023}.
Theory predicts that to grow such massive SMBHs within the first Gyr of cosmic time, they must inhabit rare, biased dark matter peaks ($\rm{M_{halo}}>10^{12.5}\,\rm{M}_\odot$), which should consequently be surrounded by abundant galaxy populations \citep[e.g.,][]{Mo2002, Volonteri2012, Costa2014, DiMascia2021, Zana2022}.
While quasars reside in rich environments on average, the source-to-source scatter is substantial, ranging from strongly overdense regions \citep[e.g.,][]{Mignoli2020, Overzier2022} to average or underdense fields \citep[e.g.,][]{Izumi2018, Eilers2024, Wang2026}.
This scatter leaves it debated whether the rapid growth of early SMBHs requires a sustained gas supply fed by large-scale structures \citep[e.g.,][]{Costa2014, DiMascia2021} or is primarily driven by internal host-galaxy processes and local gas dynamics \citep[e.g.,][]{Habouzit2022, Pizzati2024}.

\begin{figure*}[]
    \includegraphics[width=\linewidth]{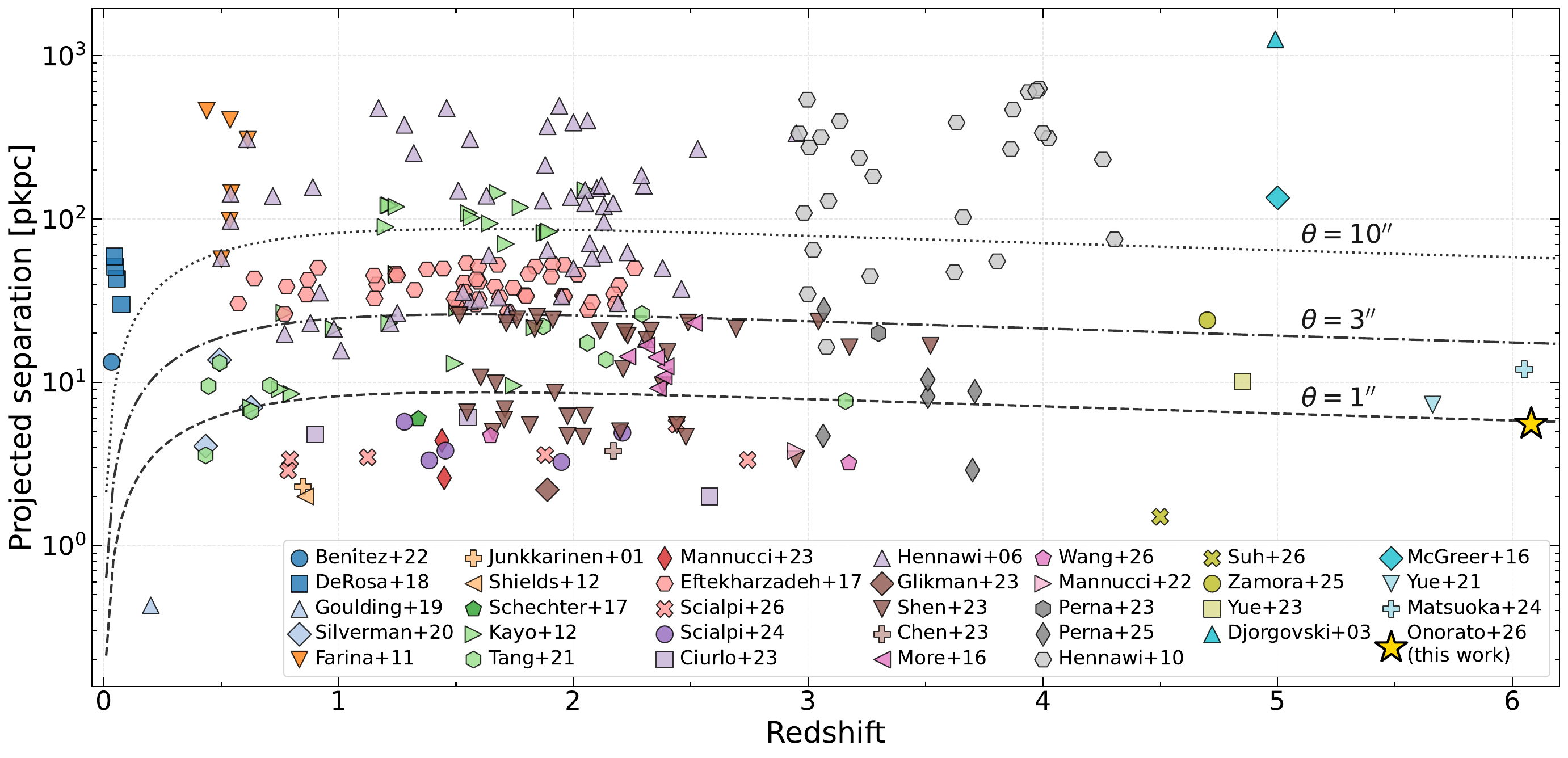}
    \centering
    \caption{Projected separation of the main literature quasar pairs and \textit{N\=ap\=owaw\=a`enakaulua} (this work) sorted by increasing mean redshift of the sample: \citet{Benitez2022, DeRosa2018a, Goulding2019, Silverman2020, Farina2011, Junkkarinen2001, Shields2012, Schechter2017, Kayo2012, Tang2021, Mannucci2023, Eftekharzadeh2017, Scialpi2026, Scialpi2024, Ciurlo2023, Hennawi2006, Glikman2023, Shen2023, Chen2023, More2016, Wang2026a, Mannucci2022, Perna2023, Perna2025, Hennawi2010, Suh2026, Zamora2025, Yue2023, Djorgovski2003, McGreer2016, Yue2021, Matsuoka2024}. The dashed, dashed-dotted, and dotted lines indicate constant angular separations of $1^{\prime\prime}$, $3^{\prime\prime}$, and $10^{\prime\prime}$, respectively.}
    \label{fig:pairs}
\end{figure*}

Pairs of quasars separated by only a few proper kiloparsecs (pkpc) in the hierarchical cosmological context are likely the signposts of extreme peaks in the matter density field where interactions, rapid gas inflows, and mergers can simultaneously feed star formation and accretion of the two SMBHs at the same time \citep[e.g.,][]{Dimatteo2005, Volonteri2022}.
At intermediate-$z$ ($2 \lesssim z \lesssim 4$), close quasar pairs are known to inhabit richer environments than single quasars \citep[e.g.,][]{Onoue2018}, making them high-yield targets for uncovering nascent protoclusters and studying SMBH fueling.
The small-scale dynamics of these systems remain largely unexplored.
At $<10$ pkpc, SMBHs in merging galaxies are expected to spiral down to a few-pc scales via dynamical friction \citep[e.g.,][]{Merritt2013}, creating bound binaries that eventually coalesce, as suggested by the recent \texttt{NANOGrav} results \citep{Agazie2023}. Currently, only a handful of quasar pairs at $z>1$ have been found at kpc-scale separations.
From the Sloan Digital Sky Survey (SDSS), \cite{Kayo2012} identified just four pairs at $z<2$ with $\sim1^{\prime\prime}$ separation ($\sim8$ pkpc; Fig. \ref{fig:pairs}). This small sample already challenges the traditional halo occupation model, which appears to underpredict the clustering signal at the smallest separations. Using \textit{Gaia}, \citet{Mannucci2022, Mannucci2023} and \citet{Shen2023} showed that the fraction of quasar pairs further increases at $\sim0.3^{\prime\prime}$ separation ($\sim3$ pkpc; Fig. \ref{fig:pairs}), implying a relevant role of direct interactions in triggering quasar activity.

While hundreds of binary quasars at separations greater than $1^{\prime\prime}$ have been discovered to date \citep[e.g.,][]{Hennawi2006, Eftekharzadeh2017}, only a few are at $z>3$ \citep[e.g.,][]{Hennawi2010, Yue2023}, with even fewer at separations less than $1^{\prime\prime}$ \citep{Shen2023}, and just three known at $z >5$ (see Fig. \ref{fig:pairs}).
\citet{McGreer2016} found the first bright quasar pair at the end of the Epoch of Reionization, CFHTLS J0221$–$0342 at $z=5$ (projected separation of $21^{\prime\prime}$; i.e., $135$ pkpc and $\Delta v \approx 150$ km s$^{-1}$).
More recently, J2037$-$4537 at $z=5.66$ (projected separation of $1.24^{\prime\prime}$; i.e., $7.3$ pkpc and $\Delta v \approx 316$ km s$^{-1}$) was discovered by \cite{Yue2021} and J1215$-$0148 at $z=6.05$, was reported by \citet[][, projected separation of $2^{\prime\prime}$; i.e., $12$ pkpc and $\Delta v \approx 553$ km s$^{-1}$]{Matsuoka2024}. The pair nature of this last system has recently been a topic of discussion and is still unclear \citep[see][]{Matsuoka2026}.

In this work, we aim to constrain the incidence of high-$z$ quasar pairs by exploiting the \textit{James Webb Space Telescope} (JWST) GO program \#5645 \citep[dubbed the \textit{Aether survey}, PI: E. P. Farina;][, in prep.]{Farina2024}. The \textit{Aether survey} provides Near-Infrared Spectrograph \citep[NIRSpec;][]{Jakobsen2022} Integral Field Unit \citep[IFU;][]{Boker2022} observations for $200$ quasars at $5.7 < z < 7$, spanning $-28 < M_{1450} < -22$ (where $M_{1450}$ is the rest-frame absolute magnitude at $1450\,\Angstrom$; with $z_{\text{mean}}=6.12$, $M_{1450\text{,mean}}=-26.16$), and representing $\sim 43\%$ of the known quasar population in this redshift range. Designed as an unbiased, parent-sample-representative survey free from target pre-selection, \textit{Aether} aims to characterize black hole demographics, ionized outflows, host-galaxy kinematics, cool gas reservoirs, and the incidence of close companions at cosmic dawn.
Through visual inspection of all NIRSpec data cubes within this dataset, we identified a second quasar in the field of the $z=6.07$ quasar J1602+4228 \citep[discovered by][]{Fan2004}, located at virtually the same redshift with a separation of a mere $0.98^{\prime\prime}$ ($\simeq 5.56$ pkpc at $z \simeq 6.08$; Fig. \ref{fig:pairs}, Section \ref{sec:results}).

This is the only companion broad-line \textit{quasar} (defined via spectroscopically confirmed redshift and broad emission lines with full-width-at-half-maximum FWHM $>2000$ km s$^{-1}$; e.g., \citealt{Fan2023}) detected across the entire survey, while a systematic study of companion \textit{galaxies} will be presented in Decarli et al. (in prep.).
We note that our spatial search is limited by the NIRSpec IFU resolution: i.e., Point Spread Function (PSF) FWHM $\approx 0.15^{\prime\prime}$ ($\approx 0.8$ pkpc at $z \sim 6$), leaving unresolved sub-kpc pairs undetected. Additionally, a search for unresolved binaries via spectroscopic signatures (e.g., double-peaked lines) was not performed in this work.

The spectra of the two quasars reveal differences in their continuum slopes, emission line profiles (Section \ref{sec:lines}), and dust extinction levels (Section \ref{sec:reddening}). While continuum and broad-line variations can be influenced by differential microlensing, subtle differences in the narrow \oiii{} line kinematics lead us to favor a physical quasar pair scenario over gravitational lensing (Section \ref{sec:pair-lens}).
If a genuine quasar pair, J1602+4228 sits above extrapolations from large-scale clustering, implying a physical connection between the two sources (Section \ref{sec:clustering}).

The pair has been named \textit{N\=ap\=owaw\=a`enakaulua} by the A Hua He Inoa program at the `Imiloa Astronomy Center of Hawai`i, meaning ``Two celestial bodies of fathomless powerful darkness in the brilliance of heat reincarnating energy into the Universe'' (see Appendix \ref{sec:app:AHuaHeInoa}).

Throughout this Letter we assume a concordance cosmology with $H_0=70$\,km\,s$^{-1}$\,Mpc$^{-1}$, $\Omega_{\rm M}=0.3$, and $\Omega_\Lambda=1-\Omega_{\rm M}=0.7$. 
In this cosmology, at $z=6.08$, the Universe is $900$ Myr old, and an angular scale of $\theta=1^{\prime\prime}$ corresponds to a proper transverse separation of $\sim 5.67$ pkpc. Adopting \citet{Planck2020} cosmological parameters yields physical scales consistent with our assumed cosmology within a few percent, having a negligible effect on our derived quantities and conclusions. The reported magnitudes are in the standard AB photometric system \citep{Oke1974, Oke1983}.

\section{Observations and Data Reduction}\label{sec:observations}
\subsection{JWST/NIRSpec IFU}\label{sec:jwst}
Observations of \textit{N\=ap\=owaw\=a`enakaulua} have been collected with JWST/NIRSpec in IFU mode on 2025-04-02. The instrument was configured with the G395H high-resolution grating and the F290LP long-pass filter, providing $3^{\prime\prime} \times 3^{\prime\prime}$ spectral cubes with resolution of $R \sim 2700$ over the wavelength range $2.87-5.27 \mu$m. The data were acquired using the \texttt{NRSIRS2} readout pattern with nine groups and one integration per exposure. To ensure optimal spatial sampling and mitigate detector artifacts, we employed a four-point \texttt{SMALL CYCLING} dither pattern, resulting in a total effective exposure time of $\sim 45$ minutes.

The data were processed using the official JWST Science Calibration Pipeline\footnote{\url{https://jwst-pipeline.readthedocs.io}.} (version 1.17.1) with the Calibration Reference Data System (CRDS). The reduction followed the standard three-stage process \citep[see][; Decarli et al. in prep., for a more detailed description of the data reduction steps]{Loiacono2024, Decarli2024}, including detector-level corrections, the assembly of individual exposures into calibrated three-dimensional (3D) data cubes, and the final combination of dithered exposures into a single rectified cube with a spatial scale of $0.1^{\prime\prime}/$pixel.

We extracted the 1D spectra of the quasars from the final combined cube using $3$ pixel circular apertures ($0.3^{\prime\prime}$) centered on the source's peak emission.
The local background was estimated and subtracted using the sigma-clipped median flux within a centered annulus with an inner and outer radius of $10$ and $15$ pixels ($1.0^{\prime\prime}$ and $1.5^{\prime\prime}$), respectively.
To account for wavelength-dependent flux losses due to the fixed extraction aperture size, we applied an aperture correction as in \citet{Loiacono2024}.
Since the JWST/NIRSpec IFU data of the \textit{Aether} program have been obtained without target acquisition, we re-aligned the NIRSpec IFU cube astrometry with archival \textit{Hubble Space Telescope} (HST) images (GO program \#13356, PI: I. McGreer; which will be presented in a subsequent work, i.e., Onorato et al., in prep.). This required applying an angular correction of $\sim 0.03^{\prime\prime}$ in R.A. and $\sim 0.10^{\prime\prime}$ in Dec.
The final JWST/NIRSpec IFU field-of-view (FoV) and spectra for the two quasars are shown in Fig. \ref{fig:spectra}.

\subsection{Gemini/GNIRS}\label{sec:gnirs}
Complementary data of \textit{N\=ap\=owaw\=a`enakaulua} have been acquired using the Gemini Near-Infrared Spectrograph \citep[GNIRS;][]{Elias2006a, Elias2006b} on the Gemini North telescope using the cross dispersion prism, the $32$ l/mm grating, and the $0.68^{\prime\prime}$ slit.
This provides wavelength coverage of the \emph{YJHK} bands ($0.81-2.52\,\mu$m) at a resolution of $R=\lambda/\Delta \lambda \simeq 750$.
The data were compiled from two separate programs: archival observations from the GN-2017A-LP-7 Large Program \citep[PI: Y.\ Shen;][, with data collected at the average parallactic-angle targeting only the brighter quasar in the pair]{Shen2019} and new dedicated observations from GN-2025A-DD-107 (PI: E.\ P.\ Farina, with the slit oriented to include both quasars). All observations were executed using a standard ABBA dither sequence with individual exposure times of $300$ s, for a total on-source exposure of $18000$ s for the bright source (combining the two programs) and $12000$ s for the fainter one.

The GNIRS data were reduced using the open-source Python-based Spectroscopic Data Reduction Pipeline \textsc{PypeIt} \citep[v1.18.1;][]{Prochaska2020, Prochaska2020a}. The pipeline performed standard image processing, including flat-fielding, automated echelle order tracing, and wavelength calibration using sky OH lines. Cosmic rays were removed using the L.\ A.\ COSMIC algorithm \citep{vanDokkum2001}, and sky subtraction was achieved via a combination of A-B differencing and B-spline fitting \citep{Bochanski2009}.
The 1D spectrum of the bright component was generated using optimal extraction \citep{Horne1986}.
An alternative approach was instead adopted for the extraction of the faint quasar, due to its faintness and proximity to the brighter companion. First, the \textsc{PypeIt}-generated object trace model of the bright source, without local sky subtraction (to avoid artificial flux suppression of the faint source), was subtracted from each A-B reduced 2D spectrum. Then, these model-subtracted A-B output frames were 2D coadded. The resulting 2D spectrum clearly revealed the faint object trace, enabling successful automatic detection and extraction.
Flux calibration and telluric absorption correction were performed using sensitivity functions from spectroscopic standard stars and a principal-component-analysis-based grid modeling approach \citep{Davies2018b}, respectively.
To account for slit losses due to the fixed position angle configuration \citep{Filippenko1982} for the spectra collected as part of the GN-2025A-DD-107 program, we applied a wavelength-dependent corrective curve (see Appendix \ref{sec:ARC}).
Absolute flux calibration was obtained by rescaling the spectra to archival \emph{J} band photometry.
We point the reader to \cite{Onorato2025} and Appendices \ref{sec:ARC} and \ref{sec:fluxscale} for a more detailed description of these data reduction steps.
The final GNIRS spectra for the two quasars are shown in Fig. \ref{fig:spectra}.

\subsection{Gemini/GMOS}\label{sec:gmos}
To complement the near-infrared (NIR) spectrum of the faint companion, we obtained optical spectroscopy using the Gemini Multi-Object Spectrograph \citep[GMOS-N;][]{Hook2004} on the Gemini North telescope (Program ID: GN-2026A-DD-105; PI: S.\ Onorato). Observations were carried out using the \texttt{R150\_G5308} grating combined with the \texttt{GG455\_G0305} order-blocking filter and a $1.0^{\prime\prime}$ slit. This setup covers the wavelength range $0.46-1.03$ $\mu$m with a resolution of $R \simeq 315$. The data were collected in $2\times2$ binning mode across $10$ individual exposures of $1200$ s, yielding a total integration time of $12000$ s.

The GMOS data were reduced using \textsc{PypeIt} following the procedure outlined in Section \ref{sec:gnirs}, with a dedicated approach to isolate the spectrum of the faint quasar from a nearby contaminant. We first ran \texttt{pypeit\_coadd\_2dspec} by placing a manual extraction aperture on the spatial trace of the foreground galaxy, enabling \textsc{PypeIt} to model and extract its 2D spatial-spectral profile directly from the coadded frame. This galaxy model was then subtracted from the 2D coadded science image, effectively removing its flux contribution. Subsequently, a new manual extraction aperture was applied to the decontaminated trace of the quasar, producing a clean 1D spectrum. Flux calibration and telluric corrections were performed using standard procedures identical to those described in Section \ref{sec:gnirs}.

We combined the GMOS optical spectrum with the parallactic-corrected GNIRS NIR spectrum of the faint quasar using the \texttt{multi\_combspec} \textsc{PypeIt} routine \citep[see][]{Onorato2025}. This routine operates on the calibrated 1D spectrum files from both instruments, producing a unified spectrum spanning [6000, 25200] $\Angstrom$ resampled to a common velocity pixel scale of $dv_{\rm{pix}}\simeq 141$ km s$^{-1}$ (coarser than GMOS spectral pixel scale). Finally, we applied absolute flux scaling by anchoring the final coadded spectrum to the synthetic \emph{K} band magnitude derived from the GNIRS data (Appendices \ref{sec:ARC} and \ref{sec:fluxscale}). The final combined GMOS+GNIRS spectrum of the faint quasar is presented in Fig. \ref{fig:spectra}.

\begin{table*}
\centering
    \caption{Properties of the \textit{N\=ap\=owaw\=a`enakaulua} quasar pair, where `A' corresponds to the bright component and `B' to the faint one.}
    \label{tab:prop}
    \begin{tabular}{lccc}
        \hline
        Properties & `A' & `B' & Note\\
        \hline
        R.A. [J2000] & $16^h 02^m 53.98^s$ & $16^h 02^m 53.91^s$\\
        Dec. [J2000] & $+42^{\circ} 28^{\prime} 24.88^{\prime\prime}$ & $+42^{\circ} 28^{\prime} 25.58^{\prime\prime}$\\ \hline
        $M_{1450}$ & $-26.92 \pm 0.02$ & $-23.74 \pm 0.72$ & $M_{1450\text{,B}}$ is a first-order estimate (Appendix \ref{sec:fluxscale}).\\
        $E(B-V)$ (continuum) & $0.014^{+0.005}_{-0.001}$ & $0.075^{+0.005}_{-0.001}$ & Intrinsic: after correcting for Galactic extinction.\\
        $E(B-V)$ (Balmer) & $0.214 \pm 0.024$ & $1.074 \pm 0.152$ & From the Balmer decrement.\\ \hline
        Projected separation & \multicolumn{2}{c}{$(0.98\pm0.01)^{\prime\prime}\approx 5.56 \pm 0.06$ pkpc} \\ 
        $\Delta v_{\rm sys}$ [km s$^{-1}$] & \multicolumn{2}{c}{$15 \pm 12$}\\
        $z_{\rm sys}$ & $6.0781 \pm 0.0001$ & $6.0785 \pm 0.0003$\\ \hline
        Continuum slope ($\alpha$) & $-1.866 \pm 0.003$ & $-0.971_{-0.015}^{+0.016}$ \& $-2.316_{-0.018}^{+0.020}$ & $\lambda_{\rm{break}}=6000 \, \Angstrom$ in the rest-frame, for `B'. \\
        FWHM$_{\text{\hb{}}}$ [km s$^{-1}$] & $3677^{+44}_{-46}$ & $2721^{+470}_{-378}$ & \\
        FWHM$_{\text{[O III],broad}}$ [km s$^{-1}$] & $1661^{+27}_{-29}$ & $1696^{+291}_{-161}$ \\
        FWHM$_{\text{[O III],narrow}}$ [km s$^{-1}$] & $537\pm7$ & $314^{+73}_{-39}$ \\
        FWHM$_{\text{\ha{}}}$ [km s$^{-1}$] & $3664^{+14}_{-13}$ & $3206^{+100}_{-96}$ & \\
        $L_{5100}$ [erg s$^{-1}$] & $(2.556 \pm 0.002)\times 10^{46}$ & $(5.803 \pm 0.013)\times 10^{45}$\\
        $L_{\text{\hb{}}}$ [erg s$^{-1}$] & $(4.439^{+0.080}_{-0.077})\times 10^{44}$ & $(8.7^{+1.5}_{-1.3})\times 10^{42}$ \\
        $L_{\text{\ha{}}}$ [erg s$^{-1}$] & $(1.733^{+0.033}_{-0.035})\times 10^{45}$ & $(8.54^{+0.23}_{-0.22})\times 10^{43}$ \\ \hline
        $L_{\text{bol}}$ [erg s$^{-1}$] & $(2.367 \pm 0.002)\times 10^{47}$ & $(5.374 \pm 0.012)\times 10^{46}$\\
        $M_{\text{BH,\hb{}}}$ [$M_\odot$] & $(1.530^{+0.036}_{-0.038})\times 10^9$ & $(3.99^{+1.65}_{-0.97})\times 10^8$\\
        $M_{\text{BH,\ha{}}}$ [$M_\odot$] & $(1.754 \pm 0.024)\times 10^9$ & $(2.55^{+0.18}_{-0.16})\times 10^8$\\
        $\lambda_{\rm{Edd,\hb{}}}$ & $1.228^{+0.029}_{-0.030}$ & $1.07^{+0.37}_{-0.30}$\\
        \hline
    \end{tabular}
    \tablecomments{Quoted uncertainties on spectral parameters, $L$, $M_{\text{BH}}$, and $\lambda_{\text{Edd}}$ represent formal $1\sigma$ statistical errors from spectral fitting. Intrinsic systematic uncertainties are larger: $\sim 0.3$ dex for $M_{\text{BH}}$ \citep{Shen2024}, $\sim 0.2$ dex for $L_{\text{bol}}$ \citep{Runnoe2012}, and $\sim 0.36$ dex for $\lambda_{\text{Edd}}$.}
\end{table*}

\subsection{Line Fitting}\label{sec:lines}
To extract the physical properties from the JWST/NIRSpec 1D spectra of the two quasars, we performed a multi-component spectral decomposition using the \textsc{python} package \textsc{Sculptor} \citep{Schindler2022}. The procedure for the full sample will be described in Protu\v{s}ov{\'a} et al. in prep., while here we report the most relevant information.

We modeled the quasar continuum using a two-stage approach. First, the power-law continuum and the \feii{} emission were fitted separately to provide initial guesses. For the bright quasar, we adopted a single power-law ($F_\lambda \propto \lambda^\alpha$) with slope $\alpha = -1.866 \pm 0.003$. For the faint source, the Bayesian Information Criterion \citep[BIC;][]{Schwarz1978} favored a broken power-law transitioning at $6000\,\Angstrom$ rest-frame, yielding slopes of $\alpha = -0.971_{-0.015}^{+0.016}$ and $-2.316_{-0.018}^{+0.020}$ in the \hb{} and \ha{} spectral regions, respectively.
Subsequently, they were optimized simultaneously in several regions selected for their low contamination from other emission lines: $[4140, 4160]$, $[4435, 4770]$, $[5080, 5770]$, $[5980, 6250]$, and $[6800, 7550]\,\Angstrom$ in the rest-frame.

To model the \feii{} pseudo-continuum emission, we adopted the template from \citet{Boroson1992} for consistency with other works \citep[e.g.,][; Farina et al., in prep.]{Pan2025, Tang2026}, but we also tested the more recent \citet{Park2022} template (see Section \ref{sec:params}). Additionally, to account for the \hb{} ``red shelf'' feature, frequently observed in Active Galactic Nuclei, we included two broad Gaussian components at the positions of the \hei{} lines \citep[see][]{Veron2002}.

Following the continuum subtraction, we modeled the broad and narrow emission lines.
The broad components of \hb{} and \ha{} were modeled with multi-Gaussian profiles, with the optimal number of components determined jointly for both lines using the BIC. This analysis favored a triple-Gaussian profile for the brighter target and a double-Gaussian profile for the fainter component (with $\Delta\text{BIC} = 115$). While the \hb{} and \ha{} redshifts were tied together, their FWHM values were left unconstrained between the two transitions to account for potential optical depth gradients and differential radial structure within the Broad Line Regions (BLRs).
For multi-component emission lines, luminosities represent the sum over all broad components, with reported FWHM values measured directly from the combined composite profile.
The narrow components of H$\gamma$, H$\delta$, \hb{}, the \oiii{} doublet, \ha{} and \nii{} were tied to the same redshift and FWHM (constrained between $250$ and $1000\,\text{km}\,\text{s}^{-1}$, where the lower bound corresponds to roughly twice the instrumental resolution to avoid fitting unresolved noise features). We tied the flux ratio of both the narrow \oiii{}$\lambda\lambda\,5007,4959 \, \Angstrom$ and \nii{}$\lambda\lambda\,6583,6548 \, \Angstrom$ doublets to $3$ \citep{Dojcinovic2023}.
The \oiii{} doublet included an additional blueshifted Gaussian component to account for potential outflows and properly capture the blue wing of the profile.
To robustly estimate the uncertainties on the spectral parameters, we performed a resampling procedure; each spectrum was perturbed $1000$ times assuming a Gaussian noise distribution pixel-by-pixel with standard deviation equal to the flux uncertainty, and re-fitted.

\begin{figure*}
    \includegraphics[width=\textwidth]{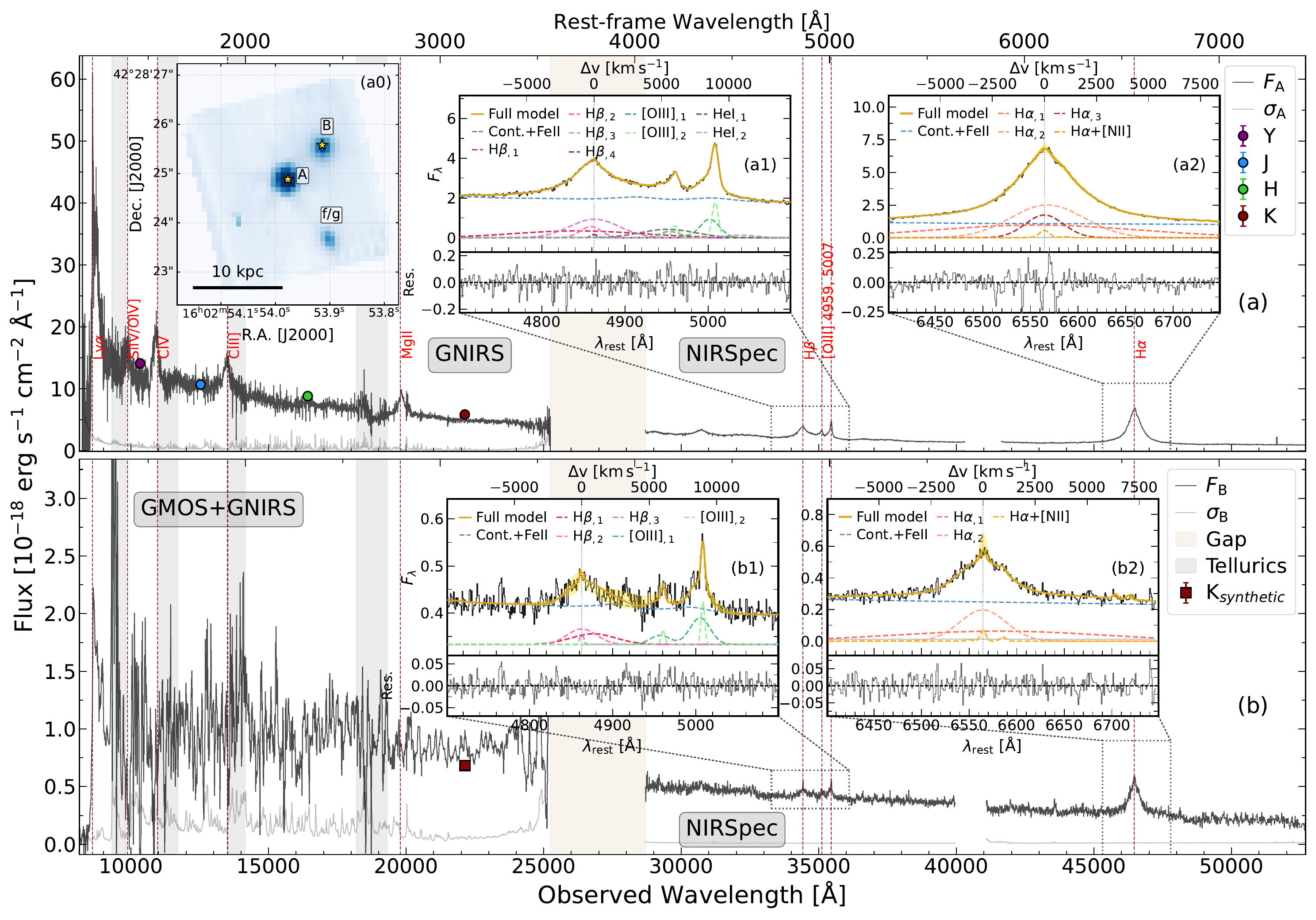}
    \centering
    \caption{\textit{(a0)}: JWST/NIRSpec IFU FoV (white light image), with the two quasars indicated by `A' and `B', marked by yellow stars, and a candidate foreground galaxy (`f/g') for which we do not identify any clear emission line.
    \textit{(a)}: Combined Gemini/GNIRS and JWST/NIRSpec spectrum of \textit{N\=ap\=owaw\=a`enakaulua} `A' (black) and associated error (gray). The light orange band highlights the gap between the spectra ([25241, 28697] $\Angstrom$). The light gray bands indicate the regions affected by telluric absorption ([13500, 14150] and [18200, 19300] $\Angstrom$). We include archival \emph{YJHK} magnitudes \citep[colored dots;][]{Ross2020}, and label the main emission lines (red dashed lines).
    \textit{(a1)}: Zoom-in on the \hb{} + \oiii{} emission line complex for `A'. Best-fit profiles are overlaid, and residuals are shown in the bottom panel.
    \textit{(a2)}: Same as \textit{(a1)}, but focusing on the \ha{} emission line.
    \textit{(b \& b1 \& b2)}: Same as in \textit{(a)}, but for the combined GMOS+GNIRS and NIRSpec spectrum of `B'. The synthetic \emph{K} band magnitude is marked by a red square and derived from the original GNIRS spectrum.}
    \label{fig:spectra}
\end{figure*}

The results of the fitting procedure for the two quasars on the \hb{} + \oiii{} and \ha{} regions are in Fig. \ref{fig:spectra}, with the relevant parameters in Table \ref{tab:prop}. Reported values are medians with uncertainties given by the $16^{\text{th}}$ and $84^{\text{th}}$ percentiles of the resampling process.

\section{Analysis and results}\label{sec:results}
We determined the location of the two quasars by collapsing the JWST/NIRSpec IFU cube along the spectral axis. Their position is reported in Table \ref{tab:prop}.
The measured angular separation is $\theta = 0.981^{\prime\prime}$. While the statistical error from the 2D Gaussian fit is small ($\sim 0.004^{\prime\prime}$), the true precision is limited by systematic uncertainties inherent to the IFU spatial sampling, cube reconstruction, and PSF asymmetries. Adopting a conservative uncertainty of $\sim0.1$ spaxels, we report an angular separation of $\theta = (0.98 \pm 0.01)^{\prime\prime}$, which resolves to a proper projected transverse distance of $5.56 \pm 0.06$ pkpc.

We established the systemic redshifts of both targets from our spectral fits (Section \ref{sec:lines}) by tying together the narrow-line components of \hb{}, \oiii{}$\lambda\lambda4959,5007\,\Angstrom$, \ha{}, and \nii{} ($z_{\rm sys} = 6.0781 \pm 0.0001$ and $6.0785 \pm 0.0003$, respectively).
We found a line-of-sight (\textit{los}) velocity separation of $\Delta v_{\rm sys} = 15 \pm 12 \text{ km s}^{-1}$. 
The \oiii{} line is on average shifted by $45$ km s$^{-1}$ with respect to systemic, with a scatter (RMS) of $\sim 100$ km s$^{-1}$ \citep[e.g.,][]{Hewett2010}. As described in \citet{Onorato2026}, we therefore conservatively adopt a fiducial uncertainty of $\Delta v_{\text{sys}} = 100$ km s$^{-1}$ \citep[see also][]{Kashino2023, Wang2023, Yang2023}.


Assuming a pure Hubble flow at $z \simeq 6.08$ (i.e., $v = H(z)\,D_{\rm proper}$, using the Hubble parameter $H(z)$ at the source redshift rather than $H_0$), this velocity offset of $\Delta v_{\text{sys}} \sim 100\text{ km s}^{-1}$ corresponds to a \textit{los} distance of $\approx 138$ pkpc. We note, however, that the observed \textit{los} velocity difference is inherently degenerate between cosmological separation and peculiar velocities. For instance, a larger physical separation could yield a small observed $\Delta v_{\text{sys}}$ if the two sources possess opposing peculiar velocities along the \textit{los}. Nevertheless, given the small projected transverse separation ($5.56 \pm 0.06$ pkpc), large 3D separations would require a highly elongated spatial configuration along our \textit{los}. While the small projected distance and velocity offset are consistent with the two quasars sharing a dense environment at cosmic dawn, spatial and velocity offsets alone cannot uniquely confirm a physical interaction. A comprehensive dynamical analysis of the gas kinematics between the two sources will be presented in Onorato et al. (in prep.).

\subsection{Physical parameters}\label{sec:params}

From the best-fit spectral models in Section \ref{sec:lines}, we derived the fundamental physical parameters of both sources. Notably, the detection of broad \ha{} and \hb{} lines (FWHM$_{\text{broad}} \gtrsim 2500$ km s$^{-1}$) supports `B' as an active type-1 broad-line quasar.
To quantify their accretion properties, we estimated the bolometric luminosity ($L_{\text{bol}}$) from the monochromatic luminosity at $5100\,\Angstrom$ ($L_{5100} \equiv \lambda L_{\lambda}$, where $\lambda=5100\,\Angstrom$) using the bolometric correction factor from \citet{Richards2006}:
\begin{equation}
L_{\text{bol}} = 9.26 \times L_{5100}.
\end{equation}
The SMBH masses ($M_{\text{BH}}$) were calculated using single-epoch virial estimators. For the \hb{} line, we adopted the most recent calibration from \citet{Shen2024}:
\begin{equation}
M_{\text{BH}} = 7 \times 10^3 \left( \frac{L_{5100}}{10^{44} \, \text{erg s}^{-1}} \right)^{0.5} \left( \frac{\text{FWHM}_{\text{H}\beta}}{10^3 \, \text{km s}^{-1}} \right)^2 M_\odot.
\end{equation}
Additionally, we derived a second mass estimate from the \ha{} line using the \citet{Greene2005} calibration\footnote{Note that the \ha{} estimator is empirically calibrated against \hb{} scaling relations and \hb{}-based reverberation mapping results, so these two estimates are not fully independent.}: 
\begin{equation}
M_{\text{BH}} = 2 \times 10^6 \left( \frac{L_{\text{H}\alpha}}{10^{42} \, \text{erg s}^{-1}} \right)^{0.55} \left( \frac{\text{FWHM}_{\text{H}\alpha}}{10^3 \, \text{km s}^{-1}} \right)^2 M_\odot.
\end{equation}
Finally, we evaluated the Eddington ratio ($\lambda_{\text{Edd}} = L_{\text{bol}}/L_{\text{Edd}}$) for both quasars by using $M_{\text{BH,\hb{}}}$ (Table \ref{tab:prop}).
Uncertainties on all derived physical quantities were computed through standard error propagation using the formal $1\sigma$ asymmetric fitting errors of line widths (FWHM) and luminosities extracted from the spectral modeling.
We note, however, that these statistical uncertainties are ultimately dominated by intrinsic systematic scatter: $\sim 0.3$ dex for single-epoch virial mass estimators \citep[e.g.,][]{Shen2024} and $\sim 0.2$ dex for bolometric corrections \citep[e.g.,][]{Runnoe2012}. Propagating these uncertainties yields a total scatter of $\sim 0.36$ dex on $\lambda_{\text{Edd}}$.

The black hole mass measurements indicate that both components host well-grown SMBHs ($M_{\text{BH}} = 1.530^{+0.036}_{-0.038}\times 10^9$ and $3.99^{+1.65}_{-0.97}\times 10^8\,M_\odot$ for `A' and `B', respectively), with values consistent between the \hb{} and \ha{} estimators within $\sim 0.06-0.20$ dex. Both quasars exhibit high accretion rates, with Eddington ratios of $\lambda_{\text{Edd}} = 1.228^{+0.029}_{-0.030}$ and $1.07^{+0.37}_{-0.30}$, respectively.

Comparing the two \feii{} templates (Section \ref{sec:lines}), the derived \hb{} parameters for component `B' are consistent within $1\sigma$ ($\Delta \log M_{\text{BH}} \approx 0.2$ dex). For component `A', the choice of template introduces a $5\sigma$ difference in $\text{FWHM}_{\text{\hb{}}}$ ($\Delta \text{FWHM} \approx 380$ km s$^{-1}$, corresponding to $\Delta \log M_{\text{BH}} \approx 0.09$ dex). This tension is driven by the small fitting uncertainties ($\sim 45$ km s$^{-1}$) enabled by the spectrum's high signal-to-noise ratio (median SNR $>100$ per $53$ km s$^{-1}$ pixel for `A', vs. $\approx 45$ for `B'). The absolute difference remains modest ($\sim 10\%$), and the properties derived from the \ha{} line, much less affected by \feii{} blending, are consistent within $0.3\sigma$ regardless of the chosen template.

To conclude, the values found are fully consistent with the properties of isolated quasars observed at similar high-$z$ \citep[e.g.,][]{Farina2022, Mazzucchelli2023, Yang2023, Belladitta2025, Liu2025, Liu2026}. Overall, even when accounting for systematic uncertainties, both quasars are accreting near or above the Eddington limit, consistent with a scenario where the ongoing interaction provides an efficient supply of gas to fuel their rapid black hole growth.

\subsection{Extinction properties}\label{sec:reddening}
A visual inspection of the spectra revealed that the fainter quasar appears redder than the brighter one. To quantify this effect, we first corrected the data for Galactic extinction using the \citet{Schlegel1998} dust maps and the \citet{Cardelli1989} extinction law ($R_V = 3.1$). Then we modeled the intrinsic reddening by applying the Small Magellanic Cloud (SMC) extinction curve \citep[$R_V = 2.7$;][]{Gordon2003} to the reference \citet{Selsing2016} quasar template, and fitting this reddened model directly to the observed spectra. This approach assumes negligible extinction from potential foreground galaxies, thus providing an upper limit on the intrinsic reddening.
The SMC law is standard for $z \gtrsim 6$ quasars, as they typically lack the $2175\,\Angstrom$ bump \citep[e.g.,][]{Hopkins2004, Gallerani2010}. We verified that the high-$z$ ``mean extinction curve'' \citep[MEC;][]{Gallerani2010} yields consistent results, but opted for the SMC law to avoid extrapolating beyond its limited rest-frame wavelength range.

Normalizing the spectra at a rest-frame wavelength of $7000\,\Angstrom$ (minimally affected by dust), we treated the color excess $E(B-V)$ as the sole free parameter, relying on the blue rest-frame coverage provided by our GNIRS (and GMOS, for `B') data to constrain the shape of the reddened continuum.
The uncertainties were derived by mapping the $\Delta\chi^2$ space around the global minimum, taking $\Delta\chi^2 = 1.0$ to define the $1\sigma$ confidence intervals.
The best-fit yields a negligible reddening of $E(B-V)_{\text{cont}} = 0.014^{+0.005}_{-0.001}$ for `A', and a higher one of $E(B-V)_{\text{cont}} = 0.075^{+0.005}_{-0.001}$ for `B'. This difference indicates distinct levels of \textit{los} obscuration between the two sources.

To independently probe the obscuration, we computed the Balmer decrement \citep[$L_{\text{\ha{}}}/L_{\text{\hb{}}}$;][]{Dong2008} from our line fits (Section \ref{sec:lines}).
Assuming a fiducial value marginally larger than the Case B recombination ratio for BLRs \citep[$\approx 3.1$, ][]{Osterbrock2006}, deviations trace \textit{los} dust extinction. The bright quasar yields $3.9\pm0.1$, which translates to $E(B-V)_{\text{Balm}} = 0.214 \pm 0.024$ (assuming an SMC extinction law). In contrast, the fainter target exhibits a heavily suppressed \hb{} flux, resulting in a decrement of $9.8\pm1.6$, corresponding to $E(B-V)_{\text{Balm}} = 1.074 \pm 0.152$.
The uncertainties on the decrements and the resulting $E(B-V)_{\text{Balm}}$ values were computed via standard analytical error propagation, utilizing the $1\sigma$ asymmetric uncertainties of the broad emission line fluxes.

While both indicators confirm that `B' is more obscured than `A', the extinction values derived from the Balmer lines are larger than those inferred from the continuum fitting. Such discrepancies are observed in quasars \citep[e.g.,][]{Dong2008, Lu2019}, reflecting the complex physical conditions of the BLR, where high optical depths and collisional excitation can alter the intrinsic Balmer ratio, as well as potential variations in the intrinsic continuum shape.
Because the observed difference in absolute magnitude between the two components is large ($\Delta M_{1450} \approx 2.89$), dust obscuration alone cannot reconcile their observed continuum flux ratios under a uniform screen model. However, the extinction measured for `B' might be boosted by a foreground galaxy almost along the same \textit{los}, detected in archival HST images (Section \ref{sec:jwst}; Onorato et al., in prep.), which could play a role when interpreting the nature of this system.

To conclude, we evaluate whether such dust attenuation could affect our derived physical quantities (Section \ref{sec:params}) by considering both reddening estimates. Correcting $L_{5100}$ (and thus $L_{\text{bol}}$) by $E(B-V)_{\text{cont}}$ yields minor increases in $L_{\text{bol}}$ by factors of $\le 1.24$ ($\lesssim 0.09$ dex). Since $M_{\text{BH,\hb{}}}$ scales with $L_{5100}^{0.5}$, the \hb{}-based masses and Eddington ratios change marginally (by $\lesssim 0.05$ dex). Conversely, correcting $L_{\text{H}\alpha}$ with $E(B-V)_{\text{Balm}}$ increases $M_{\text{BH,\ha{}}}$ by a factor of $\sim 3.26$ ($\sim 0.51$ dex) for `B'. Nevertheless, the derived accretion rates remain well within $\lambda_{\text{Edd,\hb{}}} \sim 1.2 - 1.3$. Hence, while dust attenuation is non-negligible for component `B', accounting for it does not alter our main results, reinforcing that both sources host rapidly accreting SMBHs.

\subsection{Physical quasar pair or gravitationally lensed system?}\label{sec:pair-lens}
\begin{figure*}
    \includegraphics[width=\textwidth]{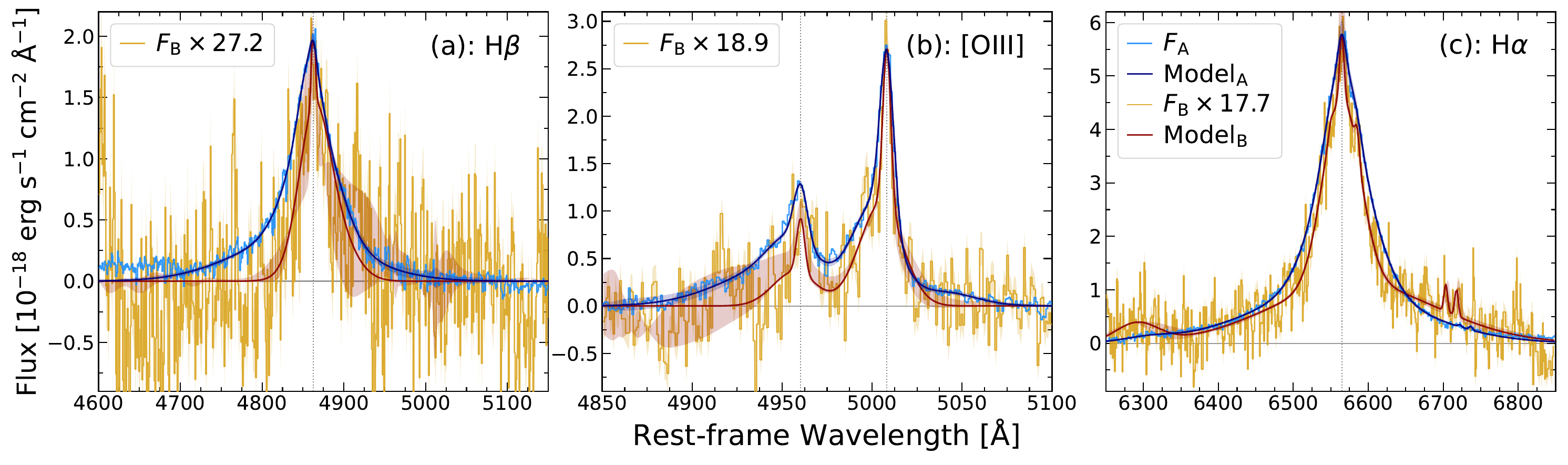}
    \centering
    \caption{Comparison of the \hb{}, \oiii{}, and \ha{} emission line profiles of \textit{N\=ap\=owaw\=a`enakaulua}, after subtracting the best-fit continuum (power law + \feii{} template) from both the data and the model, with the shaded regions showing the $1\sigma$ uncertainty.
    \textit{(a)}: \hb{} profile, obtained by additionally subtracting all \oiii{} (and \hei{}, for `A') model components.
    \textit{(b)}: \oiii{}$\lambda\lambda4959,5007\, \Angstrom$ doublet, obtained by additionally subtracting all \hb{} model components.
    \textit{(c)}: \ha{} + \nii{} blend.
    In each panel, data and best-fit model of `B' are rescaled by a multiplicative factor (quoted in the legend), so that the peak of its best-fit model matches the peak of `A' within the line of interest, allowing a direct comparison of the line profiles, despite `B' being much fainter than `A'. Vertical dotted lines mark the rest-frame wavelengths of the relevant transitions.}
    \label{fig:lines}
\end{figure*}

Given the angular separation between the two quasars of only $0.98^{\prime\prime}$ and their identical redshifts ($z \simeq 6.08$), it is essential to discuss whether the system represents a physically bound quasar pair or a gravitationally lensed quasar.

Spectral differences in the rest-frame optical continuum and in the broad emission lines (i.e., \hb{}, \ha{}; Section \ref{sec:lines}) could, in principle, be produced by differential microlensing by stars in an intervening deflector, potentially combined with intrinsic quasar variability, lensing time delays, and localized dust extinction \citep[e.g.,][]{Sluse2012, Guerras2013}.

A more stringent test of the lensing hypothesis relies on the Narrow Line Region (NLR). Because the NLR spans scales ($>100$ pc) far exceeding the Einstein radius of individual micro-lenses, its emission is immune to microlensing, making the narrow-line kinematics and profiles robust probes of the system's intrinsic nature \citep[e.g.,][]{Abajas2002, Moustakas2003, Sluse2007}. While the SNR of component `B' prevents a clean separation of the narrow components in \hb{} and \ha{}, our joint spectral decomposition of the \oiii{}$\lambda\lambda4959,5007\, \Angstrom$ doublet (panel \textit{(b)} of Fig. \ref{fig:lines}) provides crucial constraints. 

The narrow \oiii{} core components share consistent systemic velocity centroids ($\Delta v_{\rm sys} \simeq 15 \pm 12$ km s$^{-1}$), as expected for both a lensed system and a low-velocity physical pair. However, after rescaling the continuum- and broad-line-subtracted spectra, the profiles exhibit subtle differences in their blueshifted \oiii{} wings (tracing ionized outflows), with `A' displaying a more prominent blue asymmetry (see also the different FWHM$_{\text{[O III]}}$ in Table \ref{tab:prop}).
To quantify their degree of similarity, we compute a reduced $\chi^2$ statistic between the two peak-matched, continuum-subtracted profiles, using the flux uncertainties. For all three lines, we find an excess of $\chi^2_{\nu}$ over unity ($\chi^2_{\nu} = 4.0$, $4.3$, and $4.6$ for \hb{}, \oiii{}, and \ha{}, respectively), indicating that the flux rescaling does not reconcile the two profiles and their intrinsic shapes differ beyond what is expected from measurement noise alone.
While macro-lensing perturbations or differential magnification of spatially offset NLR sub-components could affect total narrow-line flux ratios \citep[e.g.,][]{Sluse2012}, differences in outflow kinematics favor two intrinsically distinct sources.

Finally, archival HST imaging (Onorato et al., in prep.) reveals a candidate foreground galaxy near the \textit{los} to `B'. This structure could host the dust responsible for the higher extinction observed in `B' ($E(B-V)_{\text{cont}} = 0.075^{+0.005}_{-0.001}$; Section \ref{sec:reddening}) and/or act as the lensing deflector. However, its relatively extended nature suggests a low-$z$, making a strong lensing scenario unlikely.
Given these competing factors, we tentatively favor the physical pair scenario while emphasizing that spectral modeling alone cannot definitively rule out lensing. A detailed lens modeling exploration will be presented in Onorato et al. (in prep.).

\subsection{Clustering}\label{sec:clustering}
\begin{figure}
    \includegraphics[width=\linewidth]{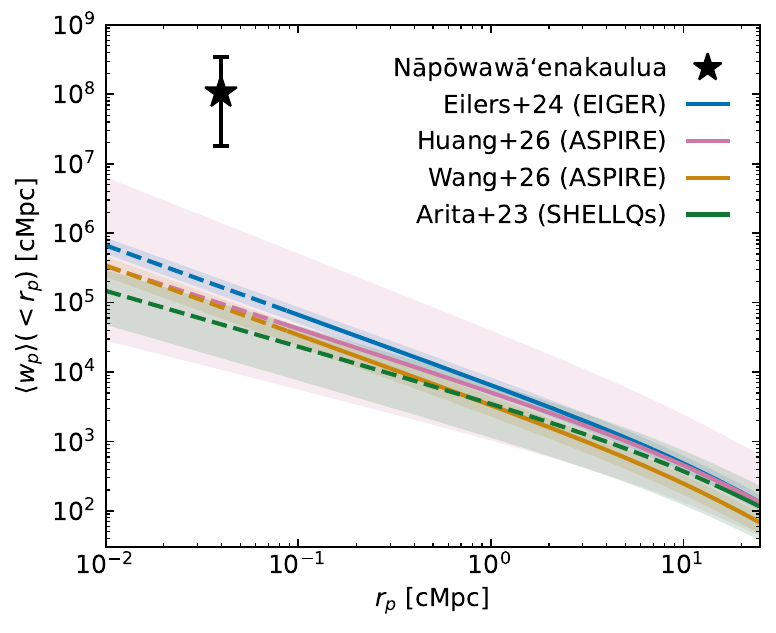}
    \centering
    \caption{Projected quasar auto-correlation function $\langle w_{p}\rangle(<r_{p})$ as a function of the projected radius $r_p$. The solid blue, pink, orange, and green lines represent the power-law fit from \citet{Eilers2024, Huang2026, Wang2026, Arita2023}, respectively, with the dashed lines being extrapolations. The black star indicates the clustering amplitude measured for \textit{N\=ap\=owaw\=a`enakaulua}.}
    \label{fig:clustering}
\end{figure}

To place this discovery in a cosmological context, we measure the small-scale clustering of high-$z$ quasars implied by the serendipitous detection of \textit{N\=ap\=owaw\=a`enakaulua}.
Around each of the $N_{\rm field}=199$ targeted quasars of the \textit{Aether survey}, we consider companion quasars of comparable luminosity within a cylinder of projected (transverse) radius $r_{p}$, equal to the observed separation of the pair, and \textit{los} half-depth $\pi_{\rm max}$. The expected number of such companions is:
\begin{equation}
\begin{split}
N_{\rm exp} &= n_{q}\!\int_{\rm cyl}\!\big[\,1+\xi(r)\,\big]\,\mathrm{d}V=\\
  &= n_{q}\,A_{\rm cyl}\, \left(2\pi_{\rm max} +\langle w_{p}\rangle(<r_{p})\,\right),
\label{eq:nexp}
\end{split}
\end{equation}
where $n_{q}$ is the blank-field number density of such quasar population, $A_{\rm cyl}=\pi r_{p}^{2}$ is the cylinder cross section, and $\langle w_{p}\rangle(<r_{p}) = A_{\rm cyl}^{-1}\!\int_{\rm cyl}\xi\, \mathrm{d}A\,\mathrm{d}\pi$ is the projected correlation function averaged over the aperture. The homogeneous (unclustered) term $n_{q}\,A_{\rm cyl}\,2\pi_{\rm max}$ is negligible here, so
every observed pair is a clustering pair. Summing over the survey and equating $N_{\rm exp}$ to the observed number of pairs $N_{\rm pair}=1$,
\begin{equation}
\langle w_{p}\rangle(<r_{p})
   = \frac{N_{\rm pair}}{N_{\rm field}\,n_{q}\,A_{\rm cyl}}\,.
\label{eq:wp}
\end{equation}
We adopt $n_{q}$ from the $z\approx6$ quasar luminosity function of \citet{Schindler2023}, integrated down to the companion magnitude $M_{1450}\approx-24$, which yields $n_{q} \simeq 9.5\times10^{-9}\,\mathrm{cMpc^{-3}}$. The aperture radius is the projected physical separation $r_{p}=5.56$ pkpc at $z=6.08$, converted to comoving units. 
With these inputs, and adopting the \citet{Gehrels1986} $1\sigma$ Poisson confidence bounds for $N_{\rm pair}=1$, Eq. \ref{eq:wp} gives:
\begin{equation}
\langle w_{p}\rangle(<r_{p})
   = 1.06^{+2.42}_{-0.88}\times10^{8}\;\mathrm{cMpc},
\label{eq:wp_value}
\end{equation}
shown in Fig. \ref{fig:clustering}.
We note that assuming $N_{\rm pair}=1$, without accounting for companion-search completeness at sub-kpc separations (Section \ref{sec:introduction}), renders our derived $\langle w_p \rangle (< r_p)$ a conservative lower limit. Furthermore, to test the impact of the uncertainty on $M_{1450\text{,B}}$ (Table \ref{tab:prop}), we recomputed $n_{q}$ using a fainter threshold of $M_{1450} < -23.0$, resulting in $\langle w_{p}\rangle(<r_{p}) = 4.81 \times 10^7\,\text{cMpc}$ (smaller by a factor of $\sim 2.2$).

We compare the value obtained in Eq. \ref{eq:wp_value} to measurements of the quasar auto-correlation function at the same redshift from the literature. 
We consider four power-law fits for the quasar auto-correlation function ($\xi_{QQ}(r)=(r/r_{0,QQ})^{-\gamma_{QQ}}$): the $z \approx 6.25$ quasar--galaxy cross-correlation function measured by the EIGER JWST program\footnote{Converted to an auto-correlation function by assuming that both quasars and galaxies trace the same linear density field.} \citep[][: $r_{0,QQ}=22.0^{+3.0}_{-2.9}\,h^{-1}$ cMpc, $\gamma_{QQ}=1.9\pm0.2$]{Eilers2024}; the analogous measurements made by the ASPIRE JWST program at $z \approx 6.6$ (\citealt{Huang2026}: $r_{0,QQ}=23.7^{+19.4}_{-9.3}\,h^{-1}$ cMpc,
$\gamma_{QQ}=1.9\pm0.2$; \citealt{Wang2026}: $r_{0,QQ}=15.8^{+2.5}_{-2.7}\,h^{-1}$ cMpc,
$\gamma_{QQ}=2.0$ fixed)\footnote{Note that the ASPIRE power-law fit by \citet{Huang2026}, compared to the one by \citet{Wang2026}, varies the slope of the auto-correlation function and includes cosmic variance in the error budget, resulting in a larger uncertainty range.}; and, on much larger scales, the quasar auto-correlation function measured by the SHELLQs program \citep[][: $r_{0}=23.7\pm11\,h^{-1}$~cMpc, $\gamma=1.8$ fixed]{Arita2023}.
To convert from $\xi_{QQ}(r)$ to $w_p(<r_p)$, we assume $\pi_{\rm max}=10$ cMpc, comparable to the integration scales of the cross-correlation analyses; this choice has little effect on the results, since $\pi_{\rm max}$ cancels in Eq. \ref{eq:wp}, and the steeply falling $\xi$ over the tiny aperture makes the model $w_{p}$ saturate by $\pi_{\rm max}\sim1$ cMpc.
We note that all of these fits are constrained on scales larger than the pair separation, and must be extrapolated to reach the pair distance; in Fig. \ref{fig:clustering} we draw each model dashed below its smallest measured bin to mark this extrapolated range.

Comparing the clustering of \textit{N\=ap\=owaw\=a`enakaulua} to these extrapolations, we find that the pair sits $\sim2-3$ orders of magnitude above the expectations, even when considering the fainter luminosity threshold. Taken at face value, this implies a very strong enhancement in the small-scale ($\lesssim5\,\mathrm{pkpc}$) clustering of quasars. Such an enhancement is partly expected, since the scales probed here lie deep in the one-halo regime; on sub-halo scales the one-halo term dominates $\xi$, and an excess of orders of magnitude over the extrapolated two-halo power-law can naturally arise. Theoretical predictions in this regime are, however, extremely uncertain. The power-law extrapolations are by construction unreliable here, and they disagree with predictions from $N$-body simulations \citep{Pizzati2024, Huang2026b}.

Nonetheless, the measured signal is so far above the extrapolations that the one-halo term alone may not fully account for it. A genuine physical connection within the pair (e.g., merger-driven triggering, which would boost the probability of simultaneous nuclear activity on these scales) could provide a further enhancement. This interpretation is appealing, but a definitive conclusion cannot be drawn from a single pair, especially given the wide variety of high-$z$ quasar environments uncovered by JWST on slightly larger scales \citep{Eilers2024, Wang2026}. Discriminating between a physically triggered small-scale enhancement and a chance close pair will require a statistical sample of such systems to constrain the one-halo regime of the quasar correlation function at these redshifts. The present system establishes that at least one such pair exists in $199$ fields, providing a first, systematics-limited anchor for that measurement.

\section{Summary and conclusions}\label{sec:summary}
In this work, we have presented the discovery and multi-wavelength characterization of \textit{N\=ap\=owaw\=a`enakaulua}, a rare, physical quasar pair at cosmic dawn ($z \simeq 6.0781$ and $6.0785$). By combining spectroscopy from JWST/NIRSpec IFU, Gemini/GNIRS and GMOS, we obtained a comprehensive view of the rest-frame UV-to-optical continuum and emission lines (\hb{} + \oiii{} complex and \ha{}) for both components. Our main findings are summarized as follows:

\begin{itemize}
    \item System properties and physical nature: the pair consists of a bright (`A': $M_{1450} = -26.92 \pm 0.02$) and a faint component (`B': $M_{1450} = -23.74 \pm 0.72$) separated by a projected distance of just $(0.98 \pm 0.01)^{\prime\prime}$ ($\approx 5.56 \pm 0.06$ pkpc) and a tight \textit{los} velocity offset of $\Delta v_{\rm sys} \simeq 15 \pm 12\,\text{km\,s}^{-1}$ (Section \ref{sec:results}). 
    \item BH demographics and accretion: multi-component spectral decomposition with \textsc{Sculptor} allowed us to derive single-epoch virial estimates of the SMBHs. For the bright component, we found $M_{\text{BH}} = (1.530^{+0.036}_{-0.038})\times 10^9\,M_\odot$, while the faint companion hosts a black hole of $M_{\text{BH}} = (3.99^{+1.65}_{-0.97})\times 10^8\,M_\odot$.
    Both systems exhibit high accretion rates ($\lambda_{\text{Edd}} = 1.228^{+0.029}_{-0.030}$ and $1.07^{+0.37}_{-0.30}$, respectively), consistent with the values of the general high-$z$ quasar population (Section \ref{sec:params}).
    \item Assessment of the gravitational lens hypothesis: while differences in continuum slopes and broad-line profiles (i.e., \hb{}, \ha{}) could be influenced by differential microlensing, the narrow \oiii{} doublet provides an insensitive probe to microlensing. The identical narrow \oiii{} velocity centroids, paired with subtle differences in their blueshifted outflow components, lead us to tentatively favor a physical pair. However, pending spectroscopic confirmation of a candidate foreground galaxy near `B', lensing cannot be fully ruled out yet (Section \ref{sec:pair-lens}).
    \item Small-scale clustering excess: we computed the projected quasar auto-correlation function $\langle w_{p}\rangle(<r_{p})$ within the volume of the parent \textit{Aether survey}. At these very close physical separations ($r \sim 5.56$ pkpc), the clustering amplitude exhibits a $\sim 2-3$ order of magnitude excess compared to the standard power-law extrapolation from large-scale studies at high redshift (Section \ref{sec:clustering}).
\end{itemize}

\appendix

\section{The meaning of \textit{N\=ap\=owaw\=a`enakaulua}} \label{sec:app:AHuaHeInoa}
Through the A Hua He Inoa program at `Imiloa Astronomy Center of Hawai'i, the following name was given to the double quasar system, \textit{N\=ap\=owaw\=a`enakaulua}. This translates to ``Two celestial bodies of fathomless powerful darkness (black holes) in the brilliance of heat reincarnating energy into the Universe'' in the indigenous Hawaiian language, `\=olelo Hawai`i.
N\=a is ``the'' for plural objects, and kaulua means ``two'' or ``double''. P\=o means ``the fathomless powerful darkness'' and refers to the central, SMBHs at the centers of the quasars. Waw\=a refers to ``reincarnation'' and `ena means ``glowing bright through heat'', which together refers to the bright radiation coming from matter accreting onto the central BHs. \textit{N\=ap\=owaw\=a`enakaulua} has connections to \textit{P\=oniu\=a`ena}, which is another high-$z$ quasar discovered by the Gemini North telescope and also contains the words p\=o and `ena \citep{Yang2020z75}. We provide the following glossary, with definitions coming from the \citet{Pukui_Elbert_1986} dictionary:

\begin{itemize}
    \item N\=a: the, plural;
    \item P\=o: powerful, fathomless darkness;
    \item Waw\=a: procreation, reincarnate \citep[see h\=anau waw\=a in the][ dictionary]{Pukui_Elbert_1986};
    \item `Ena: brightness and brilliance of heat;
    \item Kaulua: double.
\end{itemize}

\section{Atmospheric refraction correction}\label{sec:ARC}
As introduced in Section \ref{sec:gnirs}, our GNIRS observations were conducted at a fixed position angle to simultaneously observe both quasars. This setup, while necessary for the science case, introduces wavelength-dependent flux losses due to atmospheric differential refraction, as the slit is not aligned with the parallactic-angle. To recover the intrinsic flux and slope of the spectra, we developed a correction procedure using the archival spectrum of the bright component from \cite{Shen2019} as a reference.

We first flux-scaled both our GNIRS spectrum of the bright quasar and the reference \cite{Shen2019} spectrum to the \emph{H} band, as we assume that this spectral region is not significantly affected by atmospheric refraction, which is dominant only at bluer wavelengths. Then we rebinned both spectra onto a common wavelength grid with a constant velocity pixel scale ($dv_{\rm{pix}}=86.29$ km s$^{-1}$, the coarsest sampling between the two spectra), and calculated the flux ratio:
\begin{equation}
R(\lambda) = \frac{F_{\text{reference}}(\lambda)}{F_{\text{observed}}(\lambda)}.
\end{equation}
To derive a smooth correction curve, we modeled $R(\lambda)$ using a polynomial fit. To ensure the physical reliability of the fit, we implemented three key constraints:
\begin{itemize}
    \item Spectral masking: we masked regions heavily affected by telluric absorption (i.e., $[1.35,1.42]$ and $[1.82,1.93]\,\mu$m) and the Lyman-$\alpha$ forest region ($\lambda_{\text{obs}} < 1230 \times (1+z)\,\Angstrom$) to avoid biasing the fit with high-noise or absorption-dominated features.
    
    \item Anchor points: we assumed that the flux loss is minimal at longer wavelengths where the refraction effects are less pronounced. In particular, we forced the correction factor to be approximately unity in the \emph{H} and \emph{K} bands (i.e., $[1.45,1.80]$ and $[2.00,2.40]\,\mu$m, respectively) by including high-weighted ``anchor points'' in these windows.
    
    \item Model selection: to determine the optimal polynomial order and avoid overfitting, we evaluated polynomial fits of orders $2$ to $7$ using the Akaike Information Criterion \citep[AIC;][]{Akaike1974}. By minimizing the AIC, which balances goodness of fit against model complexity, we found that a $4^{\text{th}}$-order polynomial provides the statistically optimal model for our data.
    
\end{itemize}

The resulting best-fit curve represents the wavelength-dependent scaling factor required to align the fixed-angle observations with the parallactic-angle reference. Since both components of the pair were observed simultaneously in the same slit, we assume they shared the same refraction-induced losses. We therefore applied the derived correction curve to both the bright and the faint (GNIRS) quasar spectra.

\section{Absolute flux calibration, $M_{1450}$, and SNR}\label{sec:fluxscale}
The initial flux calibration of NIR spectra is typically relative, as it depends on spectroscopic standard stars that may not account for slit losses or non-photometric sky conditions. To achieve an absolute flux calibration, we anchored our reduced spectra to archival broadband photometry, as described in \cite{Onorato2025}.

For the bright component of the quasar pair, we collected archival \emph{YJHK} photometry from the UKIRT Hemisphere Survey \citep[UHS;][]{Dye2018}. We performed the absolute calibration by scaling the parallactic-corrected spectrum of the bright object to its \emph{J} band magnitude \citep[$J_{\rm AB} = 19.53 \pm 0.05$, from][]{Ross2020}. We utilized the Python library \textsc{Speclite}\footnote{\url{https://speclite.readthedocs.io/en/latest/}.} to convolve the spectrum with the UKIRT/WFCAM $J$ filter profile and derived a multiplicative scale factor ($SF$) as follows:
\begin{equation}\label{eq:scalefac}
SF = 10^{0.4 (m_{\rm spec} - m_{\rm phot})},
\end{equation}
where $m_{\rm spec}$ is the synthetic magnitude derived from the spectrum and $m_{\rm phot}$ is the photometric magnitude. For our observations, we calculated $SF=0.834$.
To assess the consistency of this calibration across the NIR range, we compared synthetic photometry from the scaled spectrum of the bright component with the available UHS magnitudes in other bands \citep[i.e., $Y_{\rm AB} = 19.65 \pm 0.06$, $H_{\rm AB} = 19.15 \pm 0.05$, and $K_{\rm AB} = 18.95 \pm 0.05$, from][]{Ross2020}. We found a fractional flux difference of $10.6\%$ in $Y$, $16.7\%$ in $H$, and $14.4\%$ in $K$.
Using the absolute flux-calibrated spectra and by following the method reported in \cite{Onorato2025}, we calculated $M_{1450} = -26.92 \pm 0.02$ for the bright quasar (in good agreement with the estimate from \citealt{Banados2016}). The uncertainty was derived by propagating the standard error of the median flux in the rest-frame $1445-1455 \, \Angstrom$ window.

A key assumption in our initial GNIRS analysis was that both components of the pair, having been observed simultaneously within the same slit, were subject to identical sky conditions.
In contrast, the GMOS data for the faint companion were acquired separately, without including quasar `A' in the slit. Therefore, the global scale factor derived from the bright component could not be applied; instead, we calibrated the faint quasar spectrum directly by anchoring it to its synthetic photometry.
Due to the significantly lower luminosity of the companion, its GNIRS spectrum is characterized by a very low SNR, particularly in the bluer NIR bands. We find median $\text{SNR}_{J,H,K} \simeq 1.3, 1.5, 3.3$ per $87.29\,\text{km\,s}^{-1}$ pixel, respectively \citep[see][ for the implemented procedure]{Onorato2025}. Given that the SNR in the \emph{J} and \emph{H} bands is near the detection limit, we consider the synthetic photometry in these windows to be dominated by noise and thus unreliable.
We only report the synthetic magnitude for the \emph{K} band, $K_{\text{AB}} \simeq 21.28 \pm 0.02$, where the signal is slightly higher. We used this \emph{K} band measurement to absolute flux calibrate the combined GMOS+GNIRS spectrum of the faint companion, from which we derived a first-order estimate of the absolute magnitude, $M_{1450} = -23.74 \pm 0.72$.
The \emph{K} band uncertainty was estimated via Monte Carlo simulations by propagating the spectral variance through the filter transmission curve.
We emphasize that these values must be treated with extreme caution. The extraction of such a faint source from GNIRS data was highly challenging, and the resulting flux is extremely sensitive to the local background subtraction and residual sky-line noise. While these estimates serve as a useful approximation of the companion's luminosity, they are subject to significantly larger uncertainties than those derived for the bright component (which hosts median $\text{SNR}_{J,H,K} \simeq 25.2, 19.4, 31.4$ per $87.29\,\text{km\,s}^{-1}$ pixel).

\section*{Acknowledgments}
The authors would like to thank the anonymous referee for the helpful comments and suggestions, which improved the clarity of the manuscript.
The authors wish to thank Ka`iu Kimura, director of the `Imiloa Astronomy Center of Hawai`i and Ka Haka `Ula O Ke`elik\=olani/College of Hawaiian Language, for her steadfast support for the A Hua He Inoa program.
This work is based in part on observations obtained at the international Gemini Observatory, a program of NSF NOIRLab, which is managed by the Association of Universities for Research in Astronomy (AURA) under a cooperative agreement with the U.S. National Science Foundation on behalf of the Gemini Observatory partnership: the U.S. National Science Foundation (United States), National Research Council (Canada), Agencia Nacional de Investigaci\'{o}n y Desarrollo (Chile), Ministerio de Ciencia, Tecnolog\'{i}a e Innovaci\'{o}n (Argentina), Minist\'{e}rio da Ci\^{e}ncia, Tecnologia, Inova\c{c}\~{o}es e Comunica\c{c}\~{o}es (Brazil), and Korea Astronomy and Space Science Institute (Republic of Korea).
The authors wish to recognize and acknowledge the very significant cultural role and reverence that the summit of Maunakea has always had within the Native Hawaiian community. We are most fortunate to have the opportunity to conduct observations from this mountain.
This work is based on observations made with the NASA/ESA/CSA James Webb Space Telescope. The data were obtained from the Mikulski Archive for Space Telescopes (MAST) at the Space Telescope Science Institute, which is operated by the Association of Universities for Research in Astronomy, Inc., under NASA contract NAS5-03127 for JWST. These observations are associated with program \#5645. The specific observations analyzed can be accessed via \href{https://archive.stsci.edu/doi/resolve/resolve.html?doi=10.17909/1ka9-hw38}{doi:10.17909/1ka9-hw38}.
S.O. acknowledges support from the JWST programs JWST-GO-4056 and JWST-GO-05645 provided by NASA through grants from the Space Telescope Science Institute, which is operated by the Association of Universities for Research in Astronomy, Inc., under NASA contract NAS5-03127.
R.D. acknowledges support from the INAF RF 2024 mini-grant  ``The interstellar medium at high redshift'' and from the PRORIS 2025 ``COSMOWEBB''.
K.P. and S.E.I.B. are supported by the Deutsche Forschungsgemeinschaft (DFG) under Emmy Noether grant No. BO 5771/1-1a.
J.-T.S. acknowledges funding by the Deutsche Forschungsgemeinschaft (DFG, German Research Foundation) - Project No. 518006966 - and support by the DFG under Germany's Excellence Strategy – EXC 2121 ``Quantum Universe'' – 390833306.
C.M. acknowledges support from Fondecyt Iniciacion grant 11240336 and the ANID BASAL project FB210003.
S.O. acknowledges Priyamvada Natarajan, Isaque Dutra, Josephine Baggen, Qian Wang, Xuheng Ding, Fabio Di Mascia, Mario Cadelano, Yuzo Ishikawa, and Dominika {\v{D}}urov{\v{c}}{\'\i}kov{\'a} for their valuable contribution to the making of this letter.

\software{\textsc{Matplotlib} \citep{Hunter2007},
        \textsc{Numpy} \citep{vanderWalt2011, Harris2020},
        \textsc{Astropy} \citep{Astropy2013, Astropy2018},
        \textsc{PypeIt} \citep{Prochaska2020, Prochaska2020a},
        \textsc{Sculptor} \citep{Schindler2022}.
          }

\bibliography{bib_highz}{}
\bibliographystyle{aasjournal}



\end{document}